\documentclass[letterpaper, 10 pt, conference]{ieeeconf} 

\usepackage{tikz}
\usetikzlibrary{shapes,shadows,arrows}
\usepackage{amsmath}
\usepackage[linesnumbered,ruled,vlined,commentsnumbered]{algorithm2e}
\usepackage{epstopdf}
\usepackage{subcaption}
\usepackage{graphicx}
\usepackage[colorinlistoftodos,prependcaption]{todonotes}
\usepackage{cite}

\newcommand{\code}[1]{\texttt{\small{#1}}}

\IEEEoverridecommandlockouts                             
\title{\LARGE \bf
Collaborative Localization and Tracking with Minimal Infrastructure
}

\author{Yanjun~Cao, David~St-Onge, Andreas~Zell and
Giovanni~Beltrame
\thanks{Dr. St-Onge, M. Cao and Dr.
Beltrame are with the Department
of Computer and Software Engineering, \'Ecole Polytechnique de
Montr\'eal, 2900 Boul \'Edouard-Montpetit, Qu\'ebec
CA e-mail: (\emph{name.surname}@polymtl.ca).}
\thanks{Prof. Zell is with the Wilhelm-Schickard-Institute for
Computer Science, University of T\"{u}bingen, Germany, e-mail:
(andreas.zell@uni-tuebingen.de).}}

\begin{document}

\maketitle
\thispagestyle{empty}
\pagestyle{empty}

\begin{abstract}
  Localization and tracking are two very active areas of research for
  robotics, automation, and the Internet-of-Things. Accurate tracking for a
  large number of devices usually requires deployment of substantial
  infrastructure (infrared tracking systems, cameras, wireless antennas,
  etc.), which is not ideal for inaccessible or protected environments. This
  paper stems from the challenge posed such environments: cover a large number of units spread over a large number of small rooms, with minimal required localization infrastructure. The idea is to accurately track the
  position of handheld devices or mobile robots, without interfering with its
  architecture. Using Ultra-Wide Band (UWB) devices, we leveraged our expertise in distributed and
  collaborative robotic systems to develop an novel solution requiring a
  minimal number of fixed anchors.  We discuss a strategy to share the UWB
  network together with an Extended Kalman filter derivation
  to collaboratively locate and track UWB-equipped devices,
  and show results from our experimental campaign tracking visitors in the Chambord castle in France. 
\end{abstract}

\section{INTRODUCTION}
Indoor localization and tracking have the potential to unlock a
plethora of new concepts and applications, both for autonomous
system research and for public space enhancement. While many
paths were explored already, it is challenging to deploy an
absolute positioning system comparable to outdoor
satellite-based GPS. Expensive and specialized setups can track
devices or people within a limited zone using laser scans or 
camera arrays, but these solutions hardly fit a large
building-wide tracking system. For applications in public
spaces, i.e., commercial halls, museums, sports centers or
schools, a localization system must be flexible, affordable and
discreet. When deployed on handheld devices, such a system can then provide the users with a more
interactive relation to the installations. On the other side of
the spectrum, technology such as RFID, can cover a large area, but
only with limited accuracy. Commercial products try to fill the
gap (GiPStech, AccuWare, Locbee, etc.) with solutions coupling
inertial measurements, geomagnetic data and available
WiFi/Bluetooth beacons or routers. These are expected to reach at
best one-meter accuracy, and quickly drift if not enough routers
are visible.

Yet the market demands for such products grows. For instance, in
museums, a localization-enabled audio-guide can help to guide the user while they navigate from room to room. It
can provide accurate contextual self-guided tours including
interactive quiz or treasure hunts, and be leveraged to share
the users experience on social networks. From the museum
operation perspective, it can provide an analytic of the floor
traffic and even help with emergency protocols.

The solution presented in this paper uses Ultra-Wide Band (UWB) network-based
measurements, able to exchange data even without line-of-sigh. In order to make our solution
relevant for most commercial human tracking applications, we aim at 10cm accuracy and try to minimize the number
of required fixed beacons (anchors). The general concept is that
each device tracked has full or partial knowledge of its own position and this
can be used to help others localize without anchors.  To ensure an optimal use of the network for distance measurements, a central monitoring system can handle the dynamics of all moving devices, but it would require full connectivity over all the building rooms and thus numerous network extenders.
However, we leverage principles from swarm
robotics based only on neighbors interactions to achieve consensus on the
UWB network usage.  Indeed, sharing a UWB network in a decentralized fashion
requires synchronization and time-sharing between the devices, so as to avoid
message collisions. To address the challenge of time-division multiple access we implement a decentralized slot assignment protocol.

The next section, Sec.~\ref{sec:rw}, introduces the research work that
inspired us in the design of our solution. Sec.~\ref{sec:sn} details our
strategy to synchronize the devices and share the network use among them. Then
Sec.~\ref{sec:tr} summarizes the measurement types and presents our
implementation of an Extended Kalman Filter.  We put everything together and
discuss the performance of the system in Sec.~\ref{sec:exp} by presenting
experiments conducted at the Chambord castle and in our laboratory. Finally, Sec.~\ref{sec:fw} describes how
the solution will be enhanced before its final large-scale deployment in the castle
scheduled in Fall 2019.

\section{RELATED WORK}
\label{sec:rw}
Our goal is to provide a easily installing global position system for visitors or robots with UWB technology. There are lots of research results using UWB sensors for localization. However, most are used to tracking one object with fixed anchors setup \cite{Benini2012, tiemann_atlas_2016, mai_local_2018, wallar_foresight_2018}. A system proposed in \cite{Prorok2014} works with multiple robots cooperative localization. However the use of UWB still belongs to fixed anchors usage and they have separate customized sensor for relative position estimation. One of the most related papers\cite{guo_ultrawideband_2017} tracking three robots is also a centralized system considering the UWB network, which almost neglects the difficulties of managing usage of UWB network. To the best of our knowledge, this paper is the first practical system using UWB technology to do localization for a group or multiple objects collaboratively in decentralized manner.

\subsection{Management of UWB network}
In order to share our UWB network without the need for a master supervision, we used two strategies: synchronization and
Time-Division Multiple Access (TDMA). The former was used in several works
focused on sensor network applications for multiple concurrent
measurements. A popular approach, the flooding-time synchronization protocol
(FTSP)\cite{marti_flooding_2004}, reaches an average time offset between
arbitrary nodes of the system. However, when considering large scattered
configurations, the nodes that require tight synchronization are usually the
closest ones. This was addressed using gradient-based
synchronization~\cite{fan2006gradient,sommer_gradient_2009}, which gives more importance to the
closest nodes to minimize the offset between clocks. To the best of our knowledge, these
techniques were never applied to the distributed usage of a UWB network.

The second body of knowledge (i.e., TDMA) includes many mechanisms to reach a
consensus on how to share network access between multiple devices in a
distributed way. Paper~\cite{xuelin_overview_2015} gives an overview of TDMA slot assignment algorithms. 
The influential algorithm Distributed Multichannel TDMA Slot Assignment Protocol (USAP)~\cite{young_usap:_1996} allows nodes to get
conflict-free slot assignment in a decentralized way using local
topology information in dynamic configurations.
This original strategy was shown to underuse the available slots, and even the extended
versions (MA)~\cite{young_usap_2000} still had to cope with a
constraint between the number of available channels and the maximum number of
neighbors considered. Furthermore, Dastango~\cite{dastangoo_performance_2009}
showed that the convergence of USAP MA is sensitive to the number of slots per
frame, requiring careful tuning from the integrator. 
A handful of further USAP extensions~\cite{kanzaki_dynamic_2003,li_evolutionary-dynamic_2007} provide a more flexible and more optimal usage of channels with adaptive frame lengths. However, the structure of their control packets became more complex, which results in heavier communication load. Chaudhary and Scheers~\cite{chaudhary_high_2012} proposed a priority mechanism allowing nodes with one-hop neighbors to be assigned slots first. While it achieved a more optimal channel usage, they did not consider the dynamic network topology. In summary, the design challenge is to get optimal channel usage while keeping low communication load.
Only two values are used in our control packets, which is even simpler than the Net Manager Operational Packet(NMOP) used in the original USAP~\cite{young_usap:_1996}.

\subsection{Localization by sensor fusion}
Works in mobile robotics have shown reliable performance of the UWB/IMU
combination for indoor positioning of rovers~\cite{Benini2012} and
quadcopters~\cite{Mueller2015}. However, both focused on a single robot tracking and used a centralized UWB setup, synchronized over Ethernet.


Benini et al.~\cite{Benini2012} considered the IMU as the process input and
they derived a non-linear process noise including bias. Merging the Ubisense
(UWB) measurement with a low-cost IMU, they showed that the localization
accuracy can be improved. Mueller et al.~\cite{Mueller2015} used the IMU to
estimate the drag force of quadcopters and input this measure into their EKF,
together with measurements from a custom UWB radio. These examples demonstrate
the high accuracy ($<10$ cm) of their strategy. For a collaborative strategy
on multi-agent localization, the work of Prorok and
Martinoli~\cite{Prorok2014} reached comparable accuracy ($<10$ cm) modeling
the UWB measurement error and compensating it with relative positioning
between the robots (a separated module based on infrared).  Instead of an EKF, they used a
particle filter, also rather commonly used with UWB. A recent study on
the collaborative use of UWB showed that two-way-ranging can give
better results than simple time-difference of
arrival~\cite{Kolakowski2016}. Their results confirmed our design choices, as previous setup
conducted with a fixed transmission scheme (not dynamic) and the tags
position computed on a central computer. Finally, the design of our EKF
was inspired by previous works that focused on tracking a single UWB tag~\cite{Hol2009,Sczyslo2008}.

\section{SHARING THE NETWORK}
\label{sec:sn}
\begin{figure}[!h]
	\begin{center}
		\includegraphics[scale=0.9]{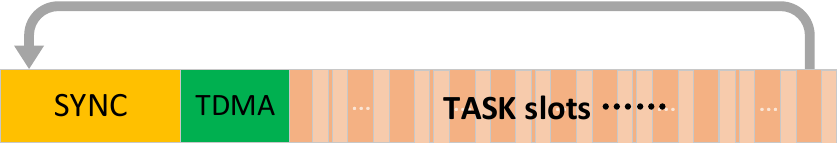}
		\caption{Transmission cycle, including synchronization, slot allocation (TDMA) and slot execution (TASK).}
		\label{fig:workflow} 
	\end{center}
\vspace{-2em}
\end{figure}

While current UWB solutions can achieve the accuracy required
for interactive museum devices, all the current commercial
products are centralized. They require either a tight
synchronization between the anchors (direct time-of-flight
measurements), done on a separated common network, or the
allocation of network slots through a master node. In both
cases, the deployment requires full connectivity of the tags and
anchors at all time. However, to limit the number of anchors and
other network-related devices (router, repeater, etc.), each tag
should be able to adapt its network usage collaborating with its
current neighbors. We achieved this by first synchronizing all
nearby tags and then splitting a cycle in slots attributed
automatically to each tag, in our case audio-guides or other user
devices to be tracked. The whole transmission mechanism runs in
cycles, as shown in Fig.~\ref{fig:workflow}: first the
synchronization, then the slots attribution, and finally each
slot execution. Each phase of the cycle uses specific custom
serialized messages transmitted over the UWB network.

\subsection{Synchronization}
Our scenario has two characteristics influencing the selection of a
synchronization algorithm: 1. a highly dynamic environment and 2. its local
network usage. The system is meant to be deployed in a museum each day,
meaning that a single user may be visiting alone, or hundreds simultaneously,
moving from room to room, thus quickly changing the network topology. A group
of people in the same room will be fully connected, while the topology will
include multiple hops and eventually be disconnected when moving
away. It would be rather difficult to keep all visitors synchronized. In any
case, localization uses neighboring anchors and tags, so nodes that are
far away can be considered less important for synchronization. A gradient-based
synchronization solution is well suited to this context.


From~\cite{fan2006gradient}, each tag uses its hardware clock:

\begin{equation}
H_i(t) = \int_{t_0}^{t} h_i(\alpha)
d\alpha+\phi_i(t_0)
\end{equation}

where $h_i(\alpha) $ is the clock rate at time $t$ and $\phi_i(t_0)$ its offset at time $t_0$.
This hardware clock should not be directly adjusted, so we define a logical clock
instead:
\begin{equation}
L_i(t) = \int_{t_0}^{t} h_i(\alpha)l_i(\alpha)
d\alpha+\theta_i(t_0)
\end{equation} 
where $l_i$ is the relative rate compensating for the drift of
the logical clock relative to the hardware clock and
$\theta_i(t_0)$, the logical clock offset. The compensation rate
$l$ can be set to 1, if we postulate that our hardware clocks
are sufficiently accurate (relative to our scenario). This way, we
consider only a single offset to maintain a synchronized
logical clock.
The objective is for each tag to maintain this logical clock as close as
possible to its neighbors'. The messages sent in this phase of the transmission
cycle, \code{syn msg}, contains a single piece of information: the logical
clock value when the message was sent. Each tag receives these messages from
all neighbors $N_i$, so it can adjust its own clock by computing the average
offset:

\begin{equation}
\label{equ:off}
\theta_i(t_{k+1})=\frac{\sum_{j\in N_i}
(L_j(t_k)-L_i(t_k)+\delta)}{|N_i|+1}\end{equation}

where $(L_j(t_k)-L_i(t_k) + \delta)$ is the clock
difference between tag $i$ and its neighbor $j$ with an
estimated communication delay $\delta$. The delays in message
transmission can be sending delay, network access delay,
propagation delay, and receiving delay. The accurate estimation
of these delays requires tedious low-level control of the
radios, but it is not mandatory to estimate the overall
$\delta$. Empirically, the average communication delay can be
estimated using simple broadcasting and receiving scripts.

In the synchronization phase, the network usage is not controlled, all tags
may access it simultaneously. To minimize collisions the tags broadcast their
\code{syn msg} randomly with a greater probability of listening to the channel
than broadcasting to it. It is important to note that some UWB systems will
hang if they experience too many packet collisions, but this strategy
generates collisions over custom messages, not using the network for
measurements. The worst case is a lost message, which can be coped with using
a sufficiently long synchronization time.
All the tags adjust their clocks towards the average of their neighbors. When
the difference between the clock and the neighbors' average is smaller than a
set threshold, or the synchronization timed out, the tags are considered
synchronized and keep their logical clock until the next cycle.


\subsection{TDMA}
Once all tags are operating in sync, a time-based schedule can be generated 
so all tags can access the UWB network in sequence for their measurements.
A TDMA algorithm includes two processes: scheduling and execution.
 In the scheduling process, all neighboring tags negotiate  which
 slots of the execution phase they can have,  converging to a consensus on the
 final sequence. To reach consensus, they exchange \code{tdma msg}, 
 serialized packets including three pieces of information: the sender id (usually provided
 by the UWB device controller), an action code,  and a requested slot id.

 If the execution phase need to be theoretically free of collision following the
 agreed sequence, the scheduling phase has to be scripted to ensure its
 performance. Following the synchronization phase, we create a neighbor table
 in each tag. This table evolves with new messages coming in and by means of a
 garbage collector.  This list is used to set an initial sequence using
 ascending IDs. This strategy assumes that a tag leaving the group 
 creates an empty slot in the schedule (not requesting any slots for the
 execution) and a new tag entering the group waits for the next round
 of scheduling.

 Every tag maintains two TDMA-related tables for scheduling: a \code{sent
   list} and a \code{received list}. From these tables, the tage can derive a
 list of free slots IDs. When broadcasting, the tag selects one of these free
 slots.  While in the listening state, the tags update their lists with the
 new messages coming in.
\begin{algorithm}
	\small
	\SetAlgoLined
	\SetKwData{Left}{left}\SetKwData{This}{this}\SetKwData{Up}{up}
\SetKwFunction{Union}{Union}\SetKwFunction{FindCompress}{FindCompress}
	\SetKwInOut{Input}{input}\SetKwInOut{Output}{output}
	\BlankLine
	\Input{ MsgReceived(actCode, slotId)}
	\Output{ SendList, RecvList.}
	$SendList, RecvList\leftarrow {-1}$\;
	$blockList \leftarrow False$\;
	\While{$len(SendList)<Threshold$}{
		\eIf{$time == myBroadcastTime$}{
			$msgSend = msgQueue(-1)$\;
			$self.broadcast(msgSend)$\;
		}{
			$msgRecv = self.listenChannel() $\;
			$(actCode,slotId)=msgRecv.decode()$\;
\uIf{$actCode == -1$ $and$ $blockList(slotId)==False$}{
$RecvList(slotId) == senderId$}
			\uElseIf{$actCode == -1$ $and$ $blockList(slotId)==True$}{
				$nextMsg = {(sender.Id,slotId)}$\;
				$msgQueue.append(nextMsg)$\;
			}
			\uElseIf{$actCode == self.Id$}{
				$SendList(slotId).disable$\;
				$nextMsg = {(self.Id,slotId)}$\;
				$msgQueue.append(nextMsg)$\;
			}
			\uElseIf{$actCode == RecvList(slotId)$}{
				$RecvList(slotId).reset$\;
			}
		}
		$blockList.update(SendList,RecvList)$
		
	}
	\caption{TDMA Schedule to reach broadcast sequence consensus}
        \label{tdma}
\end{algorithm}\DecMargin{1em}

\begin{figure}[!h]
	\begin{center}
		\includegraphics[scale=0.9]{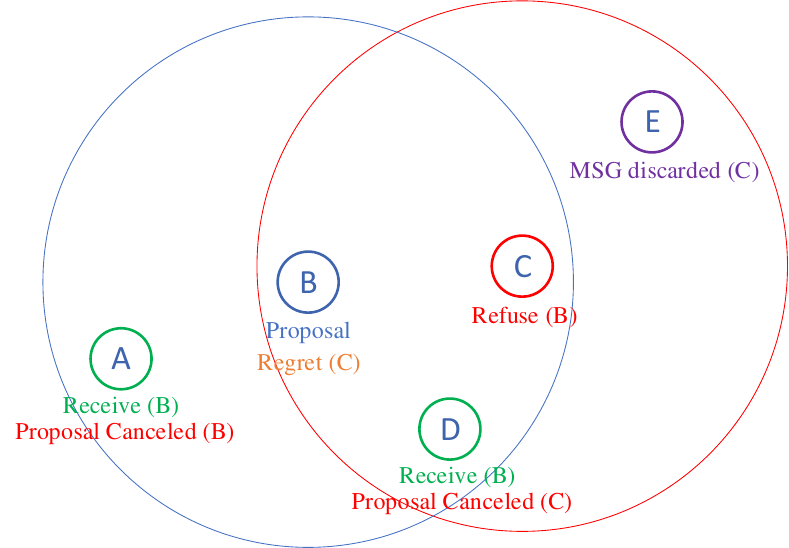}
		\caption{TDMA schedule example.}
		\label{fig:tdma01}       
	\end{center}
\vspace{-2em}
\end{figure}

Inspired by USAP, our implementation solves the TDMA schedule problem
using only broadcast messages. No peer-to-peer communication is needed, which
increases the convergence time. We introduce and example to clarify the algorithm
detailed in Alg.~\ref{tdma} and show how we avoid scheduling conflicts.
The five small circles illustrated in Fig.~\ref{fig:tdma01} are five tags, or robots. To
avoid a loaded figure, we only draw the communication ranges of tags $B$ and
$C$, which is enough to describe all cases. Tags $A$, $C$, and $D$ are within
the communication range of tags $B$. Also, tags $B$, $D$, $E$ are within the
range of tag $C$. The illustration is a snapshot of the scheduling process.

When broadcasting messages, a tag may use one of three actions: slot proposal,
proposal rejection, and proposal cancellation.  The text under the circles
represents the tag's chosen actions with the letter in parenthesis representing
the source of the incoming message.  The first tag to broadcast starts with a
slot proposal and all listeners update their \code{received list} (as tags $A$
and $D$ when receiving from $B$), as in lines 10-11 of Alg.~\ref{tdma}, unless
this slot is already flagged as attributed in their list. In this case, the
tag discovering a conflict sends a proposal rejection at its next broadcasting
slot. The tag that first sent a slot proposal may never have received a
proposal from a node with whom no direct link exists. Such a conflict with a
\emph{hidden node} is resolved with our broadcasting strategy. In
Fig.~\ref{fig:tdma01}, tag $C$ sees the conflict. A proposal rejection message
uses the conflicting tag ID as the action code. When a tag gets this type of
message, it checks if it is the source of the conflict and if it is the case,
emits a proposal cancellation message, as in lines 15-18 of
Alg.~\ref{tdma}. Otherwise, if the conflicting ID fits the one attributed in
the \code{received list}, then it knows the tag with this conflicting ID causes a conflict and
the \code{received list} entry is erased, as in lines 19-20 of
Alg.~\ref{tdma}. As showed in Fig.~\ref{fig:tdma01}, both tags $A$ and $D$
canceled the proposal of tag $B$, but following different message flows. Tag
$D$ got the proposal rejection directly from tag $C$, but $A$ is not a
neighbor of $C$, so it got only the cancellation when broadcast by $B$. This
logic prevents conflicts to emerge in the schedule, even with \emph{hidden
  nodes}.

It should be noted that for a tag that is not a neighbor of tag $B$, like $E$,
the proposal rejection message is ignored since $E$ never had a relative
entry in the received list in the first place.

\section{TRACKING THE VISITORS}
\label{sec:tr}
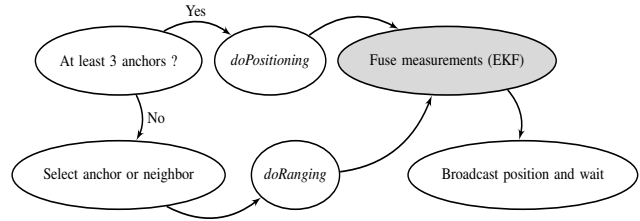
\begin{figure}[!h]
\centering
\resizebox{8.5cm}{3cm}{
\begin{tikzpicture}
\footnotesize
\begin{scope}[every node/.style={ellipse,thick,draw, inner
sep=0pt}]
\node [minimum size=12mm] (A) at (-3, 0) {At least 3 anchors ?};
\node [minimum size=12mm] (B) at (0, 0) {\emph{doPositioning}};
\node [minimum size=12mm, fill=gray!30] (D) at (3.5, 0) {Fuse
measurements (EKF)};
\node [minimum size=12mm] (E) at (5, -2) {Broadcast position and
wait};
\node [minimum size=12mm] (F) at (0.5, -2) {\emph{doRanging}};
\node [minimum size=12mm] (G) at (-3,-2) {Select anchor or
neighbor};
\end{scope}
\path [-latex',line width=.3mm] (A) edge[bend left] node[text
width=0.5cm,midway,above ] {Yes} (B);
\path [-latex',line width=.3mm] (A) edge[bend left] node[text
width=0.5cm,midway,right ] {No} (G);
\path [-latex',line width=.3mm] (B) edge[bend left]  (D);
\path [-latex',line width=.3mm] (F) edge[bend right]  (D);
\path [-latex',line width=.3mm] (D) edge[bend left]   (E);
    \path [-latex',line width=.3mm] (G) edge[bend right]  (F);
    
\end{tikzpicture}
}
\caption{The decision flow: range to a single node, and trilateration from 3
anchors or more inside a TDMA slot.}
\label{fig:decision}
\vspace{-1.5em}
\end{figure}

Our interest lies in the estimation of the real-time Cartesian position of the
devices in an absolute reference frame of the building. Which means that, for
now, we are not looking at the orientation part of the pose estimation
problem. Three different inputs are considered in the state estimation filter:
range from a neighbor, range from an anchor, and trilateration using many anchor. At each TDMA slot, the state estimation update follows
the decision flow of Fig. \ref{fig:decision}.

\subsection{Ranging}
\begin{figure}[!h]
\begin{center}
  \includegraphics[width=0.8\linewidth]{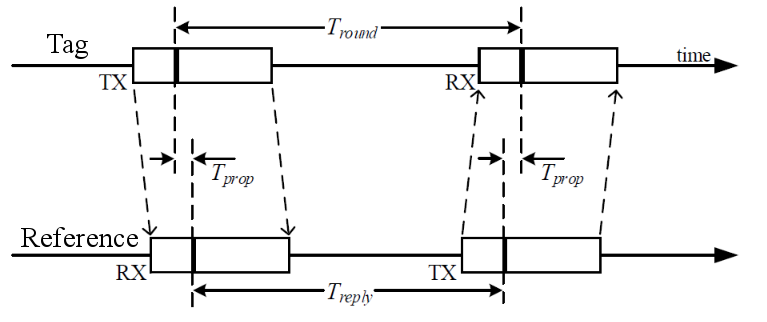}
\caption{Messages exchange while performing a single-sided
two-way-ranging.}
\label{fig:twr}       
\end{center}
\vspace{-2em}
\end{figure}

The two most common mechanisms for UWB-based localization are Two-Way-Ranging
(Pozyx, SewIO) and Time-Difference-Of-Arrival (TDoA) (BitCraze, Ubisense). While the
latter requires accurate synchronization of all the devices, it computes the
distance using a single synchronized ping from all anchors. TWR, however, uses
at least two messages, as shown in Fig. \ref{fig:twr} for the single-sided
version, and is consequently slower. The accuracy of a TWR measurement is
independent of each device's clock, thus more robust to a dynamic and large
group of tags. For instance, the TDoA mechanism of Loco (BitCraze) supports a
maximum number of 8 anchors, all in sight. To support more devices, Pozyx \cite{noauthor_pozyx_nodate}
developed a strategy exchanging the synchronization packets for all anchors
over a separated Ethernet or Wi-Fi network. This approach would require too
many communication relays in a building with walls that are impermeable to
electromagnetic waves.
The measurement becomes the average of the time difference between messages:
\begin{equation}
\text{ToF}=T_{prop}=\frac{T_{reply}-T_{round}}{2}
\end{equation}
From this measurement, the tag position can be derived, since
the distance is:
\begin{align}
d&=\text{ToF }c\\
 &=\sqrt{(x_t-x_r)^2+(y_t-y_r)^2+(z_t-z_r)^2}
\label{equ:d}
\end{align}
with $c$, the speed of light, $i_t$, the coordinate of the tag position and $i_r$, of
the reference node (anchor or neighbor).

\subsection{Trilateration}
The Pozyx system provides a filtered output of the
Cartesian coordinates of a tag. Their algorithm can work in 2D
with three anchors or in 3D with four anchors. The anchors can
be either manually entered in the code as a dictionary including
their absolute position or discovered and automatically
calibrated using ranging measurements. The second option is known
to be less accurate and so all our tests were made with manual
measurements of the anchors. The accuracy of the measurement
changes over the region covered by the anchors. We estimated an average variance of the
measurement, while keeping a tag standing still for five minutes
in three different positions in the work space and computing its
position at approximately 30Hz using four anchors:
$\mathbf{\eta}_p=\begin{bmatrix}11 & 4 & 45\end{bmatrix} cm^2$.
The anchors' position were obtained from a camera-based motion
capture (Optitrack).

\subsection{Extended Kalman Filter}
We implemented an Kalman filter to compute the three degrees of freedom in translation and their rate of change, $\mathbf{x} = \begin{bmatrix}x & \dot{x} & y & \dot{y}& z & \dot{z}\end{bmatrix}^T$ based on a changing observation vector $\mathbf{z}$. Since we expect a smooth continuous movement, the state transition matrix is a discrete time model of first order:
\begin{equation}
\mathbf{F}=\begin{bmatrix}\mathbf{D}_F & \mathbf{0} &
\mathbf{0}\\ \mathbf{0} & \mathbf{D}_F & \mathbf{0}\\ \mathbf{0}
& \mathbf{0} & \mathbf{D}_F\end{bmatrix}
\text{ with }
\mathbf{D}_F=\begin{bmatrix}1 & \delta t\\ 0 & 1\end{bmatrix} \text{ ,}
\end{equation}
and $ \mathbf{0}$ is a 2$\times$2 zero matrix, where $\delta t$ is the time elapsed since the last update of
the filter. It allows us to compute the Kalman prediction:
\begin{equation}
\hat{x}_{k\mid k-1}=\mathbf{F}\hat{x}_{k-1\mid k-1} \text{.}
\end{equation}
The prediction is then updated from the residuals vector, $\tilde{y}_k$:
\begin{equation}
\hat{x}_{k\mid k}=\hat{x}_{k\mid k-1} + \mathbf{K}_k\tilde{y}_k \text{
,}
\end{equation}
where $\mathbf{K}_k=\mathbf{P}_k\mathbf{H}_k\mathbf{S}_k^{-1}$ is the Kalman gain matrix resulting from
\begin{equation}
\mathbf {S} _{k}=\mathbf {R}_{k}+\mathbf{H}_{k}\mathbf{P}_{k\mid k-1}\mathbf{H}_{k}^T \text{ ,}
\end{equation}
and
\begin{equation}
\mathbf {P} _{k\mid k-1}=\mathbf{F}_{k}\mathbf{P}_{k-1\mid k-1}\mathbf{F}_{k}^T+\mathbf{Q}_{k} \text{ ,}
\end{equation}
with $\mathbf{R}_{k}$ and $\mathbf{H}_{k}$, respectively the observation noise and the observation matrix, varying with the available measurements. The process noise, $\mathbf{Q}_{k}$ is implemented to decrease the filter confidence on its model proportionally to the elapsed time since the last update:
\begin{equation}
\mathbf{Q}=\begin{bmatrix}\mathbf{D}_Q & \mathbf{0} &
\mathbf{0}\\ \mathbf{0} & \mathbf{D}_Q & \mathbf{0}\\ \mathbf{0}
& \mathbf{0} & \mathbf{D}_Q\end{bmatrix}
\text{ with }
\mathbf{D}_Q=\begin{bmatrix} 4\delta t & 0\\ 0 & 2\delta t\end{bmatrix}
\end{equation}
At that point the only remaining terms to defined are the observation matrix
$\mathbf{H}$ and the observation noise $\mathbf{R}$, following if the measurements
$m$ are from trilateration or range only. For trilateration, , $m = \begin{bmatrix}x & y &  z\end{bmatrix}^T$,  the three coordinates
output by the \code{doPositioning} algorithm, and
\begin{equation}
\mathbf{H}=\left[\begin{smallmatrix}1 & 0 & 0& 0 &0 &0\\
										0 & 0 & 1& 0 &0 &0 \\
										0 & 0 & 0& 0 &1 &0
\end{smallmatrix}\right] \text{ , }\mathbf{R}=\left[\begin{smallmatrix}\eta_{px} & 0& 0\\
										0 &\eta_{py} & 0\\
										0 & 0 &\eta_{pz}\end{smallmatrix}\right]
\end{equation}
The residuals vector is then
$\tilde{y}=m-\mathbf{H}^T\hat{x}$.

However, for ranging, the measurement is not linearly related to
the states (Equ.~\ref{equ:d}), thus
\begin{equation}
\mathbf{H}=\left[\begin{smallmatrix}\frac{\partial d}{\partial
x} &0 &0 & \frac{\partial d}{\partial y}& 0 &0 & \frac{\partial
d}{\partial z} &0 &0\end{smallmatrix}\right]
\end{equation}
and the residual is computed from the predicted state to
estimate the distance measured,
$\tilde{y}=d-\hat{d}$.
We compute the measurement noise matrix $\mathbf{R}$ with a confidence value for the
range input. Since the anchors are manually calibrated and their
position hardcoded in each tag, the confidence just depends on
the time passed since the measurement. However, ranging with
other visitors involves using their estimated positions, which has
an unknown level of precision. As an indicator of the
measurement accuracy, we use the condition number of the
covariance matrix $\mathbf{P}$. The ranging is noise is:
\begin{equation}
\eta_{Rt} = 
\begin{cases}
   \delta t,& \text{if } t \text{ is Anchor}\\
\log_{10}{\kappa(\mathbf{P}_t)},              & \text{otherwise}\end{cases}
\end{equation}
where
$\kappa(\mathbf{P})=\det{\mathbf{P}^{-1}}\det{\mathbf{P}}$.
Finally, since ranging measurements can be subject to wave propagation artifacts, we check the estimated distance from the predicted state of the EKF and bound the accepted readings to a maximum. If no measurements are available, for instance while waiting for the UWB network to be available, we update the EKF without measurements, but it is expected to drift slowly.

\section{EXPERIMENTS}
\label{sec:exp}

\begin{figure}[h]
\centering
\includegraphics[width=0.7\linewidth,clip,angle=180,trim={1.5cm 0.1cm 1cm 0.1cm}]{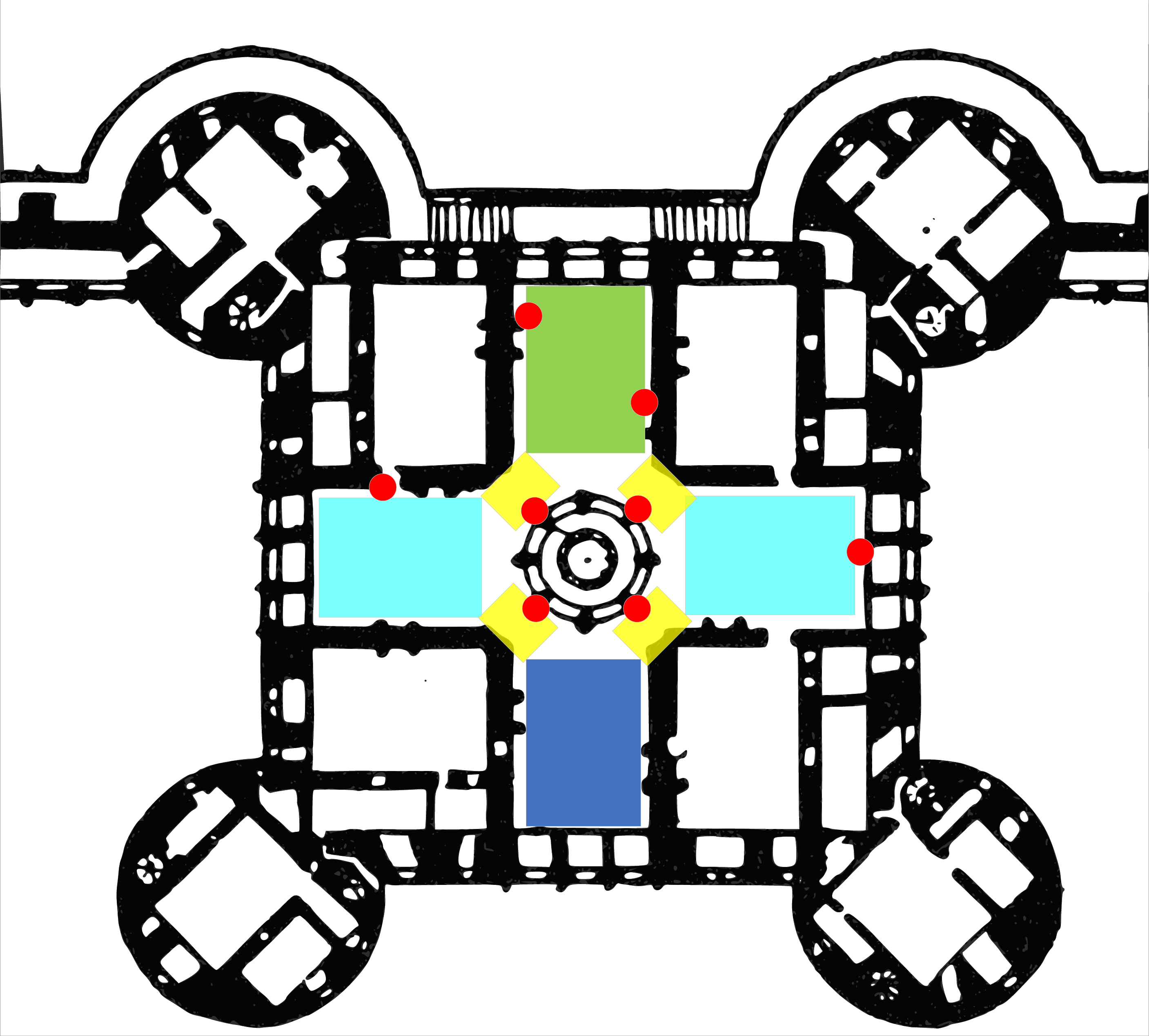}
\caption{Position of the 8 anchors (red dots) on the Chambord castle 2nd floor
  plan for the experiments. The zones color are: blue for 3 anchors, green for
  4 anchors, purple for 2 anchors, and yellow for zero anchors visible (a cornice prevented measurement from right under the anchor).
}
\label{fig:anchors}       
\end{figure}

The challenge that initiated this work is to track visitors in a museum with
only minimal infrastructure.  To test the limit of our solution, we created
different zones, as shown in Fig.~\ref{fig:anchors}: varying the number of
anchors (4, 3, 2 or 0). Seven tags, or visitors, were moving around, following
pre-defined goals in each zone, marked on the floor. We tested the
synchronization plus time management strategy to ensure no collision on the
UWB network during localization, and the accuracy of the tracking.

We started all tags together in the green zone. They first synchronized, then
coordinated. They reached a maximum time offset, with regards to their
neighbors average, of $5\ ms$. In all cases, as shown in
Fig.~\ref{fig:workflow}, synchronization occurs again at the beginning of the
next cycle, most likely with a different configuration of neighbors. The
schedule phase had slots pre-attributed following ascending IDs and until
reaching consensus on the scheduling, after which the tags wait for their slots.
 
 \begin{figure}[!h]
	\begin{center}
		\includegraphics[width=\linewidth]{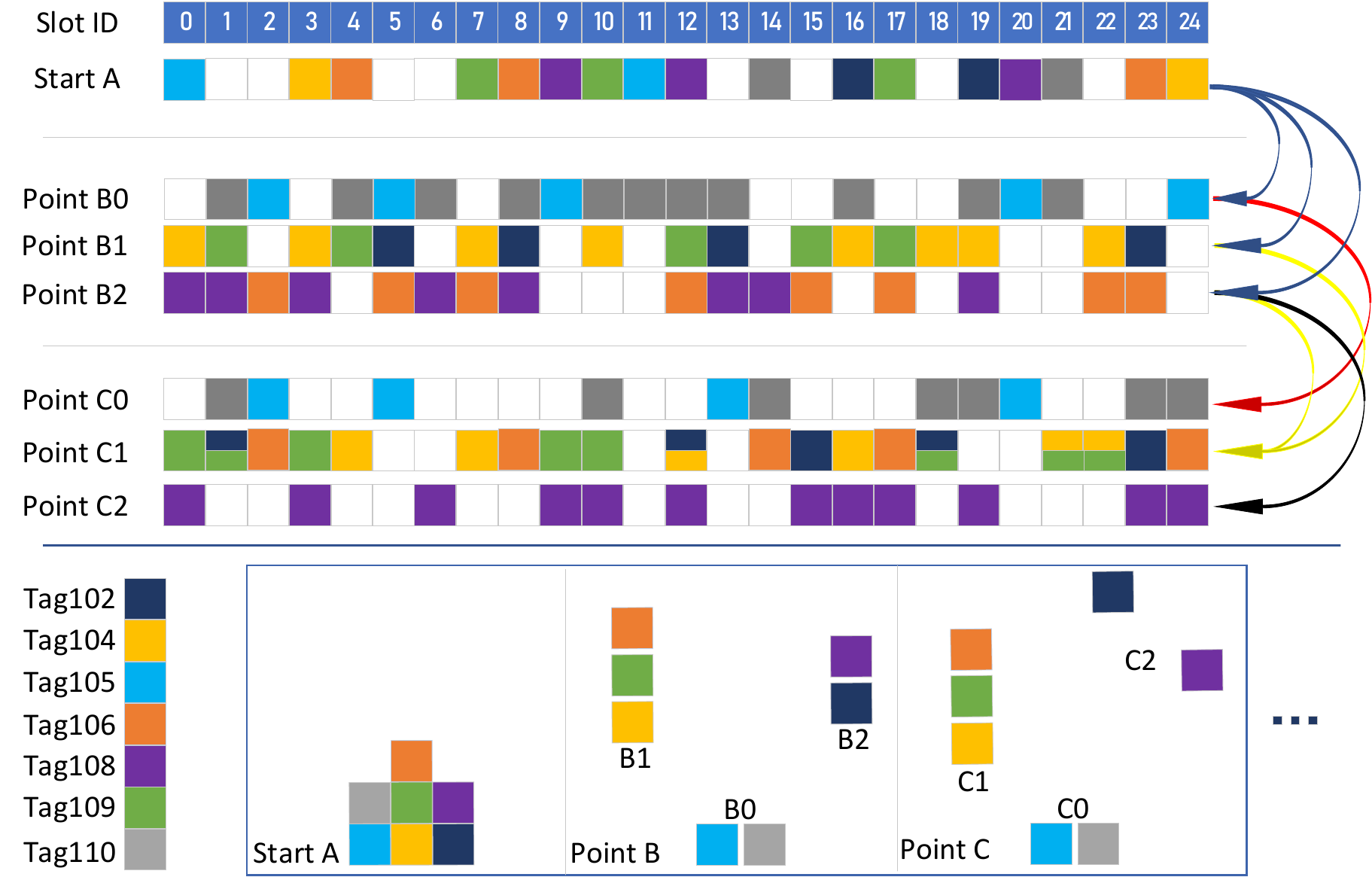}
		\caption{TDMA schedule resulting allocation tables at three
                  checkpoints. All tags started together and split in groups
                  toward a first checkpoint (Bx), then a second
                  (Cx).}
		\label{fig:tdmaresult} 
	\end{center}
	\vspace{-1.5em}
\end{figure}

Fig.~\ref{fig:tdmaresult} shows the resulting scheduling tables of one run
with the seven tags.  The number of slots available for scheduling is 25,
selected empirically to ensure all seven tags get at least a slot,
demonstrated in the first schedule table \code{Start A} of
Fig.~\ref{fig:tdmaresult}.  It should be noted that more slots do not
influence the frequency of the localization phase, since each tag will be
attributed multiple slots until the schedule is filled.  This is also visible
at all checkpoints (Bx, Cx) of Fig.~\ref{fig:tdmaresult}, in which the tags
get more slots then the initial scheduling. An interesting behavior can be
observed between the schedule of \code{Point B0} and \code{Point B1}: one
member of the B1 group received messages from Tag105 and blocked most slots
selected by this tag in group B1.  This observation relates to the
\emph{hidden node} concept discussed in Sec.~\ref{sec:sn}.

Although empty slots are visible in each schedule, the main contribution is to
ensure all possible conflicts are avoided on the UWB network.

\begin{figure}[!h]
\centering
\includegraphics[width=0.8\linewidth,trim={6.5cm 11cm 6.5cm 11cm},clip]{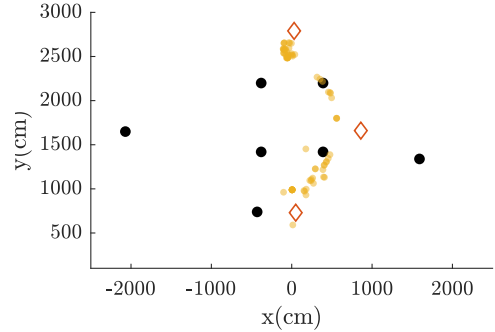}
\caption{Output positions from a visitor in the Castle (in yellow) with the anchors (in black) and the goals (in red).}
\label{fig:poseout}       
\vspace{-1em}
\end{figure}

After all tags get an approved schedule table, they start positioning
themselves, following the decision tree shown in Fig.\ref{fig:decision}. The
result of a visitor's run is shown in Fig.~\ref{fig:poseout}, walking from the
lower red diamond to the right one and finally to the top one.  Due to the few
resources (people and time) available during our experiments in the castle,
the tracking results were not as accurate as expected from our
simulations. The limitations are mainly attributed to the lack of precision on
the anchors position, as well as the precision of the ground truth (goals
marked on the ground).

\begin{figure}[!h]
\centering
\includegraphics[width=0.8\linewidth,trim={6.5cm 10cm 6.5cm 10cm},clip]{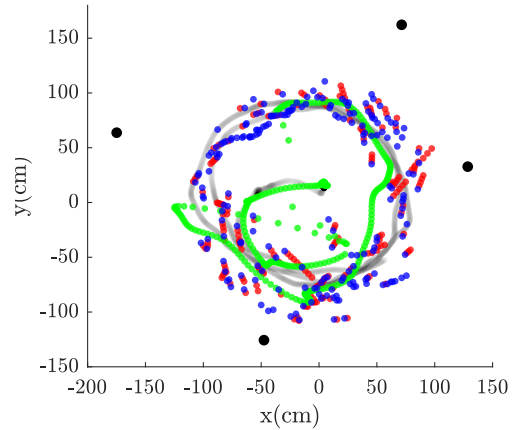}
\caption{Output positions of a person holding a tag and walking in circle. The black dots are the anchors and the gray trajectory, the ground truth from the Optitrack. The red, blue and green dots are outputs of the EKF through time following if the last update was: a trilateration (green), a range only (blue) or from the model only - no measurements (red).}
\label{fig:poseopti}       
\end{figure}

With a proper ground truth, the effect of each measurement type can be highlighted, as shown in Fig.~\ref{fig:poseopti}. Each sequence without measurements (red dots) push the EKF state to slowly drift. The trilateration measurements (green dots) are clearly the most accurate, and our range inputs stay under an acceptable average error of 9 cm.

We confirmed the two contributions of this localization strategy: none
of the tags experienced collision while conducting range measurements or
trilateration, and our estimator stays under an average error of 10 cm.

\section{FUTURE WORK}
\label{sec:fw}
Our strategy showed good performance in terms of synchronization and time
management, so it definitely allows for multiple tags to dynamically share a
common UWB network. As for the tracking system, some information, specific to
this context can be added to the workflow to help increase the accuracy. Most
visitors are walking and so using the IMU, a pedometer
approximation~\cite{Marinis2011,Jayalath2013} could give a more accurate measurement than
using the raw acceleration. Also, since the map of the building is often
available, each tag can know in which zone it is as soon as it sees an anchor
or from dead reckoning estimation. This information can help limit the drift
and remove outliers from the measurements.

These extensions will be evaluated before the final deployment, scheduled in
fall 2019.

\addtolength{\textheight}{-12cm}  


\section*{ACKNOWLEDGMENT}
\footnotesize
This project is an original idea of the NXI Gestatio [Reeves$\vert$St-Onge], a
design laboratory of the University of Quebec in Montreal. The authors would
like to acknowledge the support of the Chambord Castle team for the
experiments.


\bibliographystyle{IEEEtran}
\bibliography{swarm,hri,yanjunrefs}

\end{document}